\documentclass{apex}
\usepackage{txfonts}

\usepackage{graphicx}% Include figure files
\usepackage{dcolumn}% Align table columns on decimal point
\usepackage{bm}% bold math
\usepackage{amsmath}

\makeatletter
\newcommand*{\rom}[1]{\expandafter\@slowromancap\romannumeral #1@}
\makeatother

\title{Efficient Dual Spin-Valley Filter In Strained Silicene}

\author{Can YESILYURT$^1$, Seng Ghee TAN$^{1,2}$, Gengchiau LIANG$^1$ and Mansoor B. A. JALIL$^{1*}$}
\inst{$^1$Electrical and Computer Engineering, National University of Singapore, Singapore, 117576, Republic of Singapore \\ $^2$Data Storage Institute, Agency for Science, Technology and Research (A* Star), DSI Building, 5 Engineering Drive 1, Singapore, 117608, Republic of Singapore \\ E-mail: elembaj@nus.edu.sg}

\abst{We propose a highly efficient dual spin-valley filter in silicene, consisting of two distinct regions. In the first region, angular separation of the two valley spins in momentum-space is induced by a uniaxial strain, with further spin separation induced by an exchange field. The second region acts as an extractor of the requisite spin-valley current by means of localized fringe magnetic fields and gate modulation of the electrical potential. We demonstrated controllable and highly-efficient filtering (exceeding 90\%) for all four spin-valley combinations based on realistic parameter values. We also discussed the feasibility of practical realization of the silicene-based spin-valley filter.}

\begin{document}
\maketitle

The linear Dirac-like energy momentum dispersion of graphene has been instrumental for many spintronic and valleytronic applications\cite{1,2,3}. The utility of the Dirac dispersion has motivated a similar quest in silicene, which shares the same monolayer-honeycomb structure as graphene, but having heavier silicon atoms and consequently larger spin-orbit coupling\cite{4,5,6}.\\
In addition, silicene has the added attraction both theoretically\cite{7} and experimentally\cite{8} of real- and valley-spin dependence in its dispersion and transport properties, on account of its slightly buckled lattice. The inequivalent valleys $\ K $ and $\ K' $ at the corners of the reciprocal hexagonal lattice may be utilized in valleytronic applications. Recently, it was shown that application of strain which distorts the coupling strengths within the honeycomb lattice will also result in a valley-dependent gauge potential\cite{9,10,11}. This valley-dependent effect of strain on the electrical properties of graphene has been investigated and proposed in several application such as the quantized valley hall effect\cite{12}, modulations of I-V characteristic of nanoribbons\cite{13} and valley filtering in graphene\cite{14}. In practice, the controllable strain can be generated by depositing onto stretchable substrates\cite{15,16,17} and free suspension across trenches\cite{18}.
\par
The dependence on both the valley-spin and real-spin degrees of freedom of electron transport in silicene suggests the possibility of inducing spin-valley polarized current in the material. Such spin and valley polarization of current has been achieved magneto-optically\cite{19}. An electrical method of inducing spin-valley polarization has been proposed by means of a Y-shaped spin-valley device\cite{20}. Similarly, a spin filter based on two dimensional U-shaped device\cite{21} and a three terminal Y-shaped spin separator\cite{22} have been reported recently. Another proposal involves the generation of spin-valley polarization by means of a slicene-based lateral resonant tunneling device\cite{23}, but this is effective only on normally incident electrons. Finally, a technique\cite{24} was proposed to generate spin-valley current in silicene by application of electric field, which is however restricted to electron energy comparable to the relatively-weak spin orbit interaction energy in silicene (of the order of a few meV).
\par
In this letter, we propose a double-barrier spin-valley filter based on strained silicene, which can electrically generate highly-efficient ($\ > $ 90\%) spin-valley polarization of current. The filter relies on material and operating parameters which are accessible experimentally. Unlike previous spin-valley polarization schemes based on the spin-orbit coupling effect, our filter system is not restricted to low electron energies of the order of the spin-orbit energy split $\ \Delta _{\text{SOC}} $ (a few meV). The filtering principle is based on utilizing i) strain and exchange field to carve out distinct spin-valley transmission angular profiles in momentum-space, and ii) magneto-electric fields in the second barrier to select the particular transmission lobe corresponding to the requisite spin-valley combination. In this way, one achieves a controllable means of generating any four of the spin-valley current combinations.

%section Theory and model

We propose a two-barrier silicene filter shown in Fig. 1: Within the first barrier $\ ( 0\leq x\leq L ) $, we apply a uniform uniaxial strain on the silicene lattice, and an exchange field via adjacent magnetic insulators on top and below the silicene layer, while in the second barrier $\ ( L+a\leq x\leq 2 L+a )$, we apply a pair of $ \delta $-function magnetic fields at the boundaries $\ ( x=L+a,  \text{ and } \text{at }  x=2L+a) $ via two ferromagnetic (FM) stripes, and modulate the electrical potential via top and bottom electrostatic gates. In the model, these two regions take on different tasks: Basically, the first region is utilized to create an angular separation (i.e., in $\ \phi $-space, where $\ \phi $ is the angle of incidence) in the transmission of different spin and valley currents, while the second region is used as a barrier to sieve or filter the desired the electron current of the desired combination of spin and valley via the transverse Lorentz displacement, and block all the others. For simplicity, these two regions are analyzed separately in the analytical treatment. Subsequently, numerical analysis are carried out on both regions in an integrated manner to verify the analytical predictions.

%%%%%%%%%%%%%%%%%%%%%%%%%%%%%%%%%
%   FFFFFFFFFFFFFFFFFFFFFFFFFFFF
\begin{figure}[t!]
\begin{center}

\includegraphics[width=0.7\textwidth]{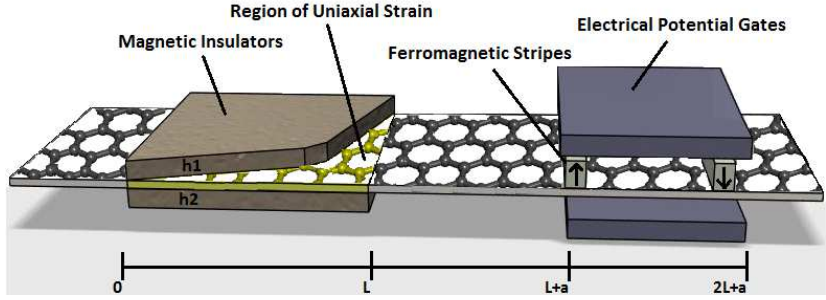}
\end{center}
\caption{Model of double-barrier, silicene based filter. Uniaxial strain is applied on the silicene region donated by yellow surface and $ h_1, h_2 $ represent magnetic insulators (e.g., EuO) yielding to spin dependent energy gap in the region of  $\ ( 0\leq x\leq L ) $. Second barrier consists of asymmetric FM stripes and electric potential barrier induced by top and bottom gates in the region of $\ ( L+a\leq x\leq 2 L+a )$. Both barrier length $L = 200$ nm and the gap between the barriers $a$ = 50 nm.}  \label{fig:ZhangJm1}
\end{figure}
%   FFFFFFFFFFFFFFFFFFFFFFFFFFFF
%%%%%%%%%%%%%%%%%%%%%%%%%%%%%%%%%

Under uniaxial strain in the $\hat{x}$ direction and lattice-dependent exchange field, the low-energy Hamiltonian can be described by

\begin{equation}
H_{\eta \sigma _z}=\text{$\hbar $v}_F \left(\hat{k}_x \tau ^x-\eta  \overrightarrow{A}_S  \hat{k}_y \tau ^y\right)+\eta \sigma _z \Delta _{\text{SOC}} +\Delta _M+\mu _{\sigma }.
\end{equation}\\

In the above,  $ \eta =\pm 1 $  denotes the valley index corresponding to valleys K and $ K' $, respectively,  while  $ \sigma _z $ is the Pauli matrix corresponding to real electron spin operator in the z-direction. As shown in Eq. 1, the intrinsic spin-orbit coupling energy term $\ \Delta _{\text{SOC}} $ couples the real and valley spin degrees of freedom, assumes a relatively small value of 3.9 meV in silicene\cite{25}. The external spin-orbit coupling (Rashba interaction) is not taken into account, as it is very small comparing to the other effects\cite{26}. The electrical potential in Eq. 1 is defined by $ \mu _{\sigma }=\mu _0+\sigma \mu _M $, where $ \mu _0 $ is the shift in the potential due, e.g., to application of the gate voltage (in the first barrier $ \mu _0=0 $ since there is no applied gate potential), while $ \mu _M $ is the contribution due to the exchange field and is given by $ \mu _M=\frac{1}{2} \left(h_1+h_2\right) $. Here, $ h_{1,2} $ refers to the exchange energy acting on the A and B sublattices of silicene, which is induced by the top and bottom magnetic insulator, respectively (Note that due to the buckling of the silicene layer, the A and B sublattice sites have a vertical separation with respect to one another). Besides the spin orbit interaction energy, also there is an additional spin-dependent effect on the band gap of system due to the exchange field which is given by $ \Delta _M=\frac{1}{2} \left(h_1-h_2\right) \sigma _z $. However, this term is zero, since we have considered the case of parallel magnetization of both the top and bottom magnetic insulators. Finally, application of strain in the $\hat{x}$ direction has opposite effect on the local hopping energy, $t\rightarrow t + \delta t~(t\approx 1.6 \text{ eV}) $ \cite{27} for bonds in the transverse  $\hat{y}$ direction. This gives rise to a strain gauge potential of $ \overrightarrow{A}_S=\delta t~\hat{y} $ which affects the $ \hat{k}_y $ component of the wave vector, thus causing a transverse deflection of the incident electrons. \\

Let us consider the expression for the electron wave functions in the three regions associated with the first barrier:

$\
\quad \quad \quad \quad \quad \quad  \Psi _\rom{1} (x)=e^{i k_x x} \left(
\begin{array}{c}
 1 \\
  M_{\eta \sigma } e^{-i \eta  \phi } \\
\end{array}\right)+\text{Re}^{-i k_x x} \left(
\begin{array}{c}
 1 \\
 -M_{\eta \sigma } e^{i \eta  \phi }  \\
\end{array}
\right);  \quad \quad \quad \quad \quad \quad x<0, \newline
\text{ } \quad  \quad \quad \quad \quad \quad \quad\Psi _{\rom{2}} (x)= A e^{i q_x x} \left(
\begin{array}{c}
 1 \\
  N_{\eta \sigma } e^{-i \eta  \theta } \\
\end{array}\right)+\text{Be}^{-i q_x x} \left(
\begin{array}{c}
 1 \\
 -N_{\eta \sigma } e^{i \eta  \theta }  \\
\end{array}
\right); \quad \quad \quad \quad \quad  0<x<L, \\
\text{ } \quad \quad \quad  \quad \quad \quad \quad \quad \quad \quad \quad \quad \Psi _{\rom{3}} (x)= T e^{i k_x x} \left(
\begin{array}{c}
 1 \\
  M_{\eta \sigma } e^{-i \eta  \phi } \\
\end{array}\right); \quad \quad  \quad \quad \quad \quad \quad \quad \quad \quad \quad x>L.
$

By substituting the above ansatze into the Hamiltonian of Eq. 1 and evaluating the corresponding eigenestates, one can evaluate $  M_{\eta \sigma } $ and $ N_{\eta \sigma } $ in the above spinor expressions. These are explicitly given by $ M_{\eta \sigma }=\frac{E+\eta \sigma \Delta _{\text{SOC}}}{\sqrt{E^2-\Delta _{\text{SOC}}^2}} $, and $ N_{\eta \sigma }=\frac{\left(E-\mu _{\sigma }\right)+\eta \sigma \Delta _{\text{SOC}}}{\sqrt{\left(E-\mu _{\sigma }\right){}^2-{(\eta \sigma \Delta _{\text{SOC}})}^2}} $. For simplicity, $ \sigma =\pm 1 $ denote spin in the $ \pm z $ direction. \\

Next, we consider the wave vectors within and outside the first barrier: The $\hat{x}$-component of these are given by $ q_x= \sqrt{\left(E-\mu _{\sigma }\right){}^2-{(\eta \sigma \Delta _{\text{SOC}}})^2}\frac{\cos (\theta )}{\text{$\hbar $v}_F} $ and $ k_x=\sqrt{E^2-\Delta _{\text{SOC}}^2}\frac{ \cos (\phi )}{\text{$\hbar $v}_F} $, respectively, where $ \theta  (\phi) $ is the angle between the electron momentum and the x-axis within (outside) the barrier. By considering the conservation of the momentum in the transverse $\hat{y}$ direction in the unstrained (I and III) and strained regions (II), we obtain the relation:

\begin{equation}
\sin (\phi ) \sqrt{E^2-\Delta _{\text{SOC}}^2} = \sin (\theta ) \sqrt{\left(E-\mu _{\sigma }\right){}^2-{(\eta \sigma \Delta _{\text{SOC}})}^2}+\eta \text{$\delta $t} ,
\end{equation}

\noindent
from which, one obtains the angle $ \theta =\sin ^{-1}\left(\frac{\sin (\phi ) \sqrt{E^2-\Delta _{\text{SOC}}^2} +\eta\text{$\delta $t} }{\sqrt{\left(E-\mu _{\sigma }\right){}^2-{(\eta \sigma \Delta _{\text{SOC}})}^2}}\right) $  within the barrier. Finally, the valley and spin dependent transmission probability $ \mathcal{T} =\left|T|^2\right. $ can be calculated by applying the wave function continuity relations at the boundaries. In Figs 2(a) and (b), it can be clearly seen that the angular profile of the transmission probability changes significantly when the strain and spin-dependent potential are varied. The dependence of $\mathcal{T}$ on the strain field $\delta $t can be explained as follows: In the presence of $\delta $t, the conservation of the transverse momentum at the interface between the unstrained and strained regions dictates that $ E \sin (\phi )=E \sin (\theta )\pm \text{$\delta $t},$ and thus, $\sin (\theta )=\sin (\phi )\mp \frac{\text{$\delta $t}}{E}$. Therefore, the change in $\delta $t directly affects the transmitted angle of electrons. Besides, as stated earlier, the momentum $q_x$ is also a function of spin-dependent potential $ \mu_\sigma $. This results in further angular separation of the transmission probability $ \mathcal{T} $ for different spin orientations.

We found that under the optimal choice of strain and exchange strength in the first barrier, the transmission of particular spin-valley combination can be compressed to within a tight range of incident angle $ \phi $. Moreover, the angles corresponding to perfect transmission $ ( \mathcal{T} \approx 1) $  for all four different spin-valley combinations are well-separated in the incident angular space $ \phi $. This is a prerequisite for highly-efficient filter operation in the second barrier region.
\par
Let us analyze the valley and spin separation within the first barrier. Based on Eq. 2, and the fact that perfect transmission in Klein tunneling occurs at normal incidence, i.e. $ \sin (\theta )=0 $, the angles corresponding to perfect transmission for the two valleys are then given by $ \phi =\eta \sin ^{-1}\left(\frac{\text{$\delta $t}}{\sqrt{E^2-\Delta _{\text{SOC}}^2}}\right). $ Hence, one observes that the separation of the $ K $ and $ K' $ valley polarization in $ \phi $-space is induced by the strain energy \text{$\delta $t}. This simple analysis applies in the limit of infinite barrier length, i.e., $ L\rightarrow \infty $. However, for a finite barrier thickness, perfect transmission $ (\mathcal{T}=1) $ occurs under the resonance condition: $ q_x L =n\pi $, where $n=0,\pm 1,\text{...}$. For a sufficiently long device length, extra resonance peaks occur corresponding to $n\ge 2$.  In addition, the dependence of $ q_x $ on the real spin due to the spin-dependent potential $ \mu _{\sigma } $ results in a further angular separation of the perfect transmission direction for opposite spins $ \sigma =\pm 1 $. These perfect transmission angles can be derived analytically for all four spin-valley combinations $ (\sigma =\pm 1, \eta =\pm 1) $ by considering both the resonance condition and the conservation of transverse momentum (Eq. 2). The analytical results agree with the location of the transmission peaks which are obtained numerically in Figs. 2(a) and (b). In the optimal case depicted in Fig. 2(a), the strain energy is chosen to be sufficiently large ($\delta t=43$ meV) so that only the first resonant peak, i.e., $ q_x L =\pm \pi $ occur, while the higher resonances correspond to angles greater than $ \frac{\pi }{2} $ and are thus totally reflected. Note that the strain energy cannot be excessively large, otherwise the transmission of even the first resonant peak will be suppressed.

%%%%%%%%%%%%%%%%%%%%%%%%%%%%%%%%
%   FFFFFFFFFFFFFFFFFFFFFFFFFFFF
\begin{figure}[t!]
\begin{center}

\includegraphics[width=0.5\textwidth]{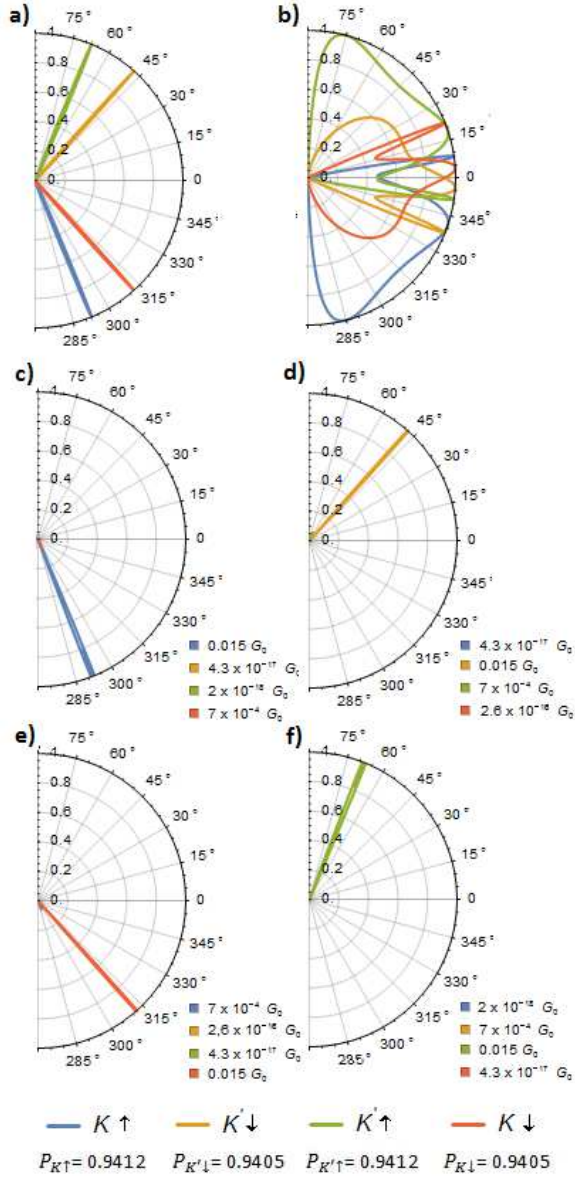}
\end{center}
\caption{The transmission probability of each spin and valley combination in terms of incident angle in the case of $ E_F $ = 25 meV, $L = 200$ nm, $a = 50$ nm, $ \Delta _{\text{SOC}} $ = 3.9 meV, a) polarization of different spin-valley in $ \phi $-space under the optimal configuration of strain and spin-dependent potential $\delta t= 43$ meV, $ h_1 = h_2 = 2 $ meV as a result of the first barrier b) angular dependence of transmission probability under non-optimal configuration of strain and spin-dependent potential $\delta t = 16$ meV, $ h_1 = h_2 = 2.5 $ meV. c), d), e) and f) show transmission probability of the system after passing through the second region. The magneto-electric potential in the second region is set at the optimal value so as to achieve an effective filter operation for all four spin-valley combinations: c) $ K_{\uparrow } $, d) $ K_{\downarrow }^\text{'} $, e) $ K_{\downarrow} $ , and f) $ K_{\uparrow }^\text{'} $. The magneto-electric configuration for the filter operations of the four spin-valley combinations are as follows: c) $\mu$ = 2 meV and $\text{$\delta $t}_B = 43$ meV, d) $\mu = -2$ meV and $\delta_B = -43$ meV, e) $\mu = -2$ meV and $\delta t_B = 43$ meV, f) $\mu =2$ meV and $\delta t_B = -43$ meV.}  \label{fig:ZhangJm1}
\end{figure}
%   FFFFFFFFFFFFFFFFFFFFFFFFFFFF
%%%%%%%%%%%%%%%%%%%%%%%%%%%%%%%%

Having achieved a distinct angular separation of the transmission peaks for all four spin-valley polarizations by passage through the first (strained) barrier, we are in a position to effectively filter the desired spin-valley polarization by means of the second barrier. The spin and valley filtering are achieved within the second barrier by applying appropriate magnetic vector and electrical potentials. Delta $ \delta $-function magnetic fringe fields $  B_z(x)=B_0 l_B[\delta  (x)-\delta  (x-L)] $ are generated at the boundaries of the second barrier (here for simplicity we set x=0 at the left boundary) by antisymmetric ferromagnetic stripes (see Fig. 1). These fields yield a square-hat vector (gauge) potential $ \overrightarrow{A}_B=B_0 l_B[\Theta (x)-\Theta  (x-L)]\hat{y} $ within the second barrier, where $ l_B=\sqrt{\frac{\hbar }{\text{eB}_0}} $ is the magnetic length. Essentially, this gauge potential induces a transverse deflection of the electrons, thus allowing a certain angular range in $ \phi $-space to transmit through the second barrier. To filter out the requisite valley polarization, $ \overrightarrow{A}_B $ should match the magnitude of the strain gauge $ \overrightarrow{A}_S $.  Physically, under this matching condition, the transverse deflection from the strain $ \overrightarrow{A}_S $ and fringe field $ \overrightarrow{A}_B $ cancels one another allowing electrons of the particular valley index $ \eta $ to transmit through. From the matching condition, $ \eta \frac{\text{$\delta $t}  }{v_F}=\text{eB}_0 l_B $, one can obtain the required magnetic fringe field strength $ B_0 l_B= \eta \frac{1}{\hbar e}(\frac{\delta t}{v_F})^2 $. In addition, to filter out the required real spin polarization, the electrical potentials of the two barrier regions should also be matched, i.e., $ \mu _{\sigma ,1}=\mu _{\sigma ,2}\rightarrow \sigma \mu _{M,1}=\mu _{0,2} $, i.e., the electrical potential $ \mu_0 $ (due to the gate bias in the second barrier) should be equal to the (exchange induced) spin-dependent potential $ \sigma \mu_M $ (in the first barrier). In this way, highly efficient filter operation can be achieved for the desired combination of spin ($ \sigma $) and valley ($ \eta $).

\par
So far, in our analytical treatment we have considered the two barrier regions separately. To verify our analytical prediction, we numerically calculate the overall transmission probability of whole system by considering the Dirac electron wave functions in all five regions and apply the proper matching conditions at the four boundaries. Subsequently, we calculate the conductance of spin-valley filter by integrating over the incident angles, i.e.,\\
\\
$\
\text{    }  \quad  \quad \quad \quad \quad \quad \quad \quad \quad \quad \quad \quad  G_{\uparrow (\downarrow )}^{K\left(K'\right)}=G_0\int _{-\frac{\pi }{2}}^{\frac{\pi }{2}}\cos (\phi )\mathcal{T}_{\uparrow (\downarrow )}^{K\left(K'\right)}d\phi $, \\
\\
where $ G_0=\frac{e^2}{h \left(E_F L_y/\text{$\hbar $v}_F \right)} $.

%%%%%%%%%%%%%%%%%%%%%%%%%%%%%%%%
%   FFFFFFFFFFFFFFFFFFFFFFFFFFFF
\begin{figure}[t!]
\begin{center}

\includegraphics[width=0.4\textwidth]{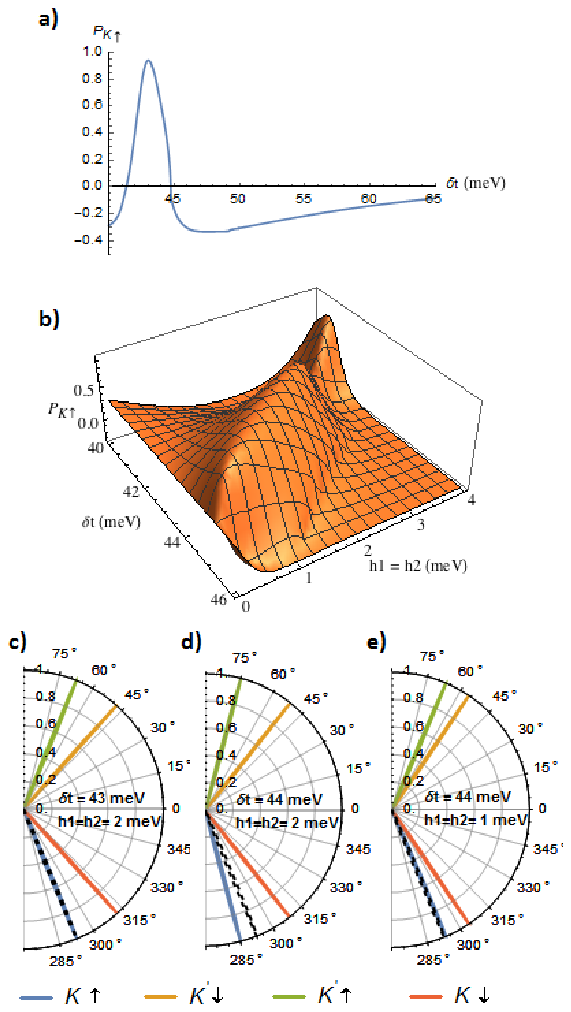}
\end{center}
\caption{The analysis of the effect of varying strain field and spin-dependent potential due to exchange field $h_{1,2}$ on the transmission $ P_{K\uparrow } $. In the following calculations, we assume the following parameter values: $ E_F $ = 25 meV, $L = 200$ nm, $a = 50$ nm, $ \Delta _{\text{SOC}} $ = 3.9 meV. (a) The polarization  $ P_{K\uparrow } $ as a function of the strain field which is varied between 40 meV and 65 meV; (b) The dependence of the polarization $ P_{K\uparrow } $ on both the strain field and exchange field $h_{1,2}$; (c), (d), and (e) show the effect of changes in the strain field and the exchange field on the angular position of the peak transmission probability across the first region. The dashed black plot represents the angular position which matches the parameters (magnetic field and electrical potential) of the second barrier (filter) region. }  \label{fig:ZhangJm1}
\end{figure}
%   FFFFFFFFFFFFFFFFFFFFFFFFFFFF
%%%%%%%%%%%%%%%%%%%%%%%%%%%%%%%%

Finally, the spin-valley polarization (e.g., for valley $ K $ and spin $ \uparrow $) is defined by \\
\\
$\
\text{}  \quad \quad \quad \quad \quad \quad \quad \quad \quad \quad \quad  P_{K\uparrow }=\frac{\left(G_{K\uparrow }-G_{K\downarrow }\right)+\left(G_{K\uparrow }-G_{K'\uparrow }\right)+\left(G_{K\uparrow }-G_{K'\downarrow }\right)}{3\left(G_{K\uparrow }+G_{K\downarrow }+G_{K'\uparrow }+G_{K'\downarrow }\right)}. $ \\

By assuming the optimal parameter values and matching conditions, we numerically obtain high valley-spin polarization $ P_{\sigma ,\eta }\geq 0.94 $  for all four spin-valley combinations. Figs. 2(c) to (f) depicts the angular dependence of transmission probability of the whole system (transmission across 5 regions) under different matching configurations. In all four cases, we obtain transmission of only the desired spin-valley polarization while the other three polarizations are virtually filtered out.  Additionally, one can control the transmitted spin-valley polarization merely by changing the sign of the magnetic and electrical potentials in the second barrier, i.e., by reversing gate-induced potential $ \mu _0\rightarrow -\mu _0 $ and the fringe magnetic field $ B_z\rightarrow -B_z $. Our numerical calculations assume a barrier width of L = 200 nm, and matching magnetic field strength of $ B_0 $ = 1 T and strain energy of \text{$\delta $t} = 25 meV. These parameter values are within reach experimentally, and thus the proposed spin-valley filter  can be realized with current technology. By contrast, the spin-valley filtering solely by means of spin-orbit coupling effect\cite{24} yields a much greater overlap of the transmission profiles for the different valleys and spins, which translates to a significantly lower polarization values. Furthermore, the valley-spin polarization exists only at low electron energies, comparable to the SOC energy of a few meV$ ' $s. Numerically, it was found that the SOC-induced valley and spin polarization attains a value of about 10\% at electron energy of 10 meV, and vanishes when the energy is increased to 20 meV.
\par
In Fig. 3(a), the valley-spin polarization $ P_{K\uparrow } $ for the K valley and spin-up electrons is plotted over the range of strain field between 40 meV and 65 meV. The plot shows a peak at 43 meV, the value which matches the magnetic field strength in the second barrier (filter) region. For values of strain $\text{$\delta $t}$ slightly above or below 43 meV, the polarization $ P_{K\uparrow } $ decreases sharply due to the mismatch of the vector potentials $ A_S $ and $ A_B $ in the first and second regions. Interestingly, when both the strain $ \text{$\delta $t} $ and the exchange field $h_{1,2}$ (and hence the spin-dependent potential) are varied, we find that the reduction in the polarization $ P_{K\uparrow } $ due to increasing strain can be counter-acted by reducing the exchange field [see Fig. 3(b)]. This behavior can be understood by considering the effect of these two parameters on the angular dependence of transmission probability. As shown in Fig. 3(c), when we set $ \text{$\delta $t} $ = 43 meV and $ h_{1,2} $ = 2 meV, the peak of the transmission curve $ \mathcal{T}_{K\uparrow} $ (blue curve) exactly coincides with the black dashed line which denotes the angular orientation which matches the second filter region. However, a small change of $ \text{$\delta $t} $ to 44 meV, results in a significant shift in the transmission in $ \phi $-space [see Fig. 3(d)] away from the dashed line. This translates into a drop in the transmitted spin-valley of $ (K,\uparrow) $ across the filter region, and hence a lower $ P_{K\uparrow } $. By modifying the exchange field to $ h_{1,2} $ = 1 meV, the peak of the $ \mathcal{T}_{K\uparrow} $  curve can be made to coincide with the dashed line [see Fig. 3(e)], and thus $ P_{K\uparrow } $ is restored to its peak value.
\par
We theoretically investigate the real and valley spin polarized transport in a double-barrier system based on strained silicene, capable of achieving high real and valley spin polarization of current (exceeding 90\%). The filtering principle is based on utilizing i) strain and exchange field in the first barrier to carve out distinct spin-valley transmission angular profiles in momentum-space, and ii) magneto-electric fields in the second barrier to select the requisite spin-valley combination and filter the others. We demonstrate, both analytically and numerically, that almost pure spin-valley current can be realized based on geometric and material parameters that are within practical feasibility.\\

\acknowledgement
We thank the MOE Tier II grant MOE2013-T2-2-125 (NUS Grant No. R-263-000-B10-112), and the National Research Foundation of Singapore under the CRP Program ''Next Generation Spin Torque Memories: From Fundamental Physics to Applications'' NRF-CRP9-2013-01 for financial support.

\end{document}